\documentclass[%
 reprint,
 amsmath,amssymb,
 aps,
]{revtex4-2}

\newcommand{\frc}[2]{\raisebox{2pt}{$#1$}\big/\raisebox{-3pt}{$#2$}}

\usepackage{graphicx}% Include figure files
\usepackage{dcolumn}% Align table columns on decimal point
\usepackage{bm}% bold math
\begin{document}

\preprint{APS/123-QED}

\title{Nonlinear Optical Effects Due to Magnetization Dynamics In a Ferromagnet}% Force line breaks with \\

\author{Evgeny A. Karashtin}
  \email{eugenk@ipmras.ru}
  \affiliation{Institute for Physics of Microstructures RAS, Nizhny Novgorod,  Russia}%Lines break automatically or can be forced with \\
  \affiliation{Lobachevsky State University of Nizhny Novgorod, Nizhny Novgorod, Russia}
\author{Tatiana V. Murzina}%
 \email{murzina@mail.ru}
\affiliation{%
  M.V. Lomonosov Moscow State University, Moscow, Russia
}%


%\affiliation{
%\affiliation{

\date{\today}% It is always \today, today,

\begin{abstract}
We theoretically consider magnetization dynamics in a ferromagnetic slab induced by the magnetic field of a strong femtosecond laser pulse. The longitudinal geometry, in which the initial magnetization lies in both the plane of incidence and the sample plane, is studied. The magnetization oscillations at the optical wave frequency are calculated with the use of the Kapitza pendulum approach taking into account that the optical frequency is much greater than the magnetization oscillation eigenfrequency. We study the reflection of the electromagnetic wave from a ferromagnet %with such a dynamics 
and show that this laser-induced low-frequency magnetization dynamics leads to the appearance of the second-order nonlinearity in the Maxwell's equations, which in turn gives rise to both the second harmonic generation (SHG) and rectification effect. Although the amplitude of the magnetization oscillations is small, the considered effect may be responsible for the SHG with the efficiency comparable to that of nonmagnetic SHG from metal surfaces. Our estimations show that the suggested mechanism may explain the recent experiments on magnetization induced modulation of the SHG intensity in a ``forbidden'' $P_{in}P_{out}$ combination of incident and reflected waves in cobalt/heavy metal systems, where it can be even more pronounced due to the spin current flow through the ferromagnet/ heavy metal interface.
%\begin{description}
%item[Structure]
%\end{description}
\end{abstract}

\maketitle

%\tableofcontents
\section{Introduction}

Nonlinear optical effects such as second harmonic generation (SHG) or rectification attract a lot of attention for the last decades. On one hand, this is governed by the fact that these effects may exist only in non-centrosymmetric systems. As a result, these nonlinear optical phenomena, and first of all the SHG probe, provide %If a medium itself is possesses the effects appear at the surface which provides 
a powerful method for studying the properties of surfaces and interfaces where the inversion symmetry is broken \cite{Shen,Aktsipetrov_JOSAB}. On the other hand, the rectification effect %provides 
is an efficient mechanism for the THz waves generation under the excitation by femtosecond laser pulses \cite{Kim_NatPhot, Gildenburg_PRL, Vvedenskii_PRL, Kampfrath_NatNano}. A special research direction here is the magnetization-induced phenomena in %solid state 
systems containing magnetic materials. For instance, THz sources based on ferromagnet~/ heavy metal multilayers are widely studied \cite{Seifert_NatPhot, Bull_APLM}. Magnetization brings new symmetry properties to a medium \cite{Pan_PRB}, which in turn leads to the appearance of the nonlinear-optical analogues of the magnetooptical Kerr and Faraday effect, and even to a number of new ones. % It is known that if a surface of a ferromagnet is irradiated by a p-polarized wave the magnetization-induced double-frequency reflected wave would be s-polarized. However, 
Among others, recently the so named ``forbidden'' magnetization-induced SHG intensity effect was observed in ferromagnet~/ heavy metal systems such as Co/Pt, Co/Ta  etc. multilayers, which consists in variation of the p-polarized SHG intensity by longitudinal dc magnetic field  \cite{Murzina_OptExpr, Kolmychek_JETP, Murzina_JETPL}. %Note that all mentioned nonlinear effects are quite small with their  efficiency being of the order of $10^{-6}$ or less. %smaller (by magnitude).

Symmetry analysis of the nonlinear-optical interactions \cite{Pan_PRB} does not take into account the effects of (expected)  magnetization dynamics induced by the electromagnetic wave. They are usually supposed  to be small 
%the magnitude of such magnetization change is 
%they are small because 
as the optical frequency exceeds substantially the eigenfrequency of magnetization oscillations.
However if a strong femtosecond optical pulse is considered with the electric field of the order of 1~MV/cm or greater, %which can be easily realized, 
the frequency ratio is of the order of $10^{-4}$, while the magnetic field of the optical wave is relatively strong.
%(may be of the order of $4 \pi M_s$ where $M_s$ is the saturation magnetization). 
Therefore one can expect that the effects that appear due to magnetization dynamics may be comparable to those  provided by static magnetization due non-linearity at a ferromagnet surface. 

In this work we study these effects and compare the theoretical results with the data of recent experiments. The magnetization dynamics caused by the magnetic field of the incident laser radiation %wave
is described in the framework of the Landau-Lifshitz-Gilbert equation solved within  the Kapitza pendulum approach. The Maxwell equations are then solved with the assumptions of small gyrotropic term of the dielectric constant and small magnetization oscillation magnitude. We obtain both double frequency electric field and zero-frequency (rectified) electric field or the electric current in a  ferromagnet
%, which depends on the conductive properties of a ferromagnet, 
induced by the electromagnetic wave. Finally, a boundary problem is solved. We suppose that the elecromagnetic wave is incident at the surface of a ferromagnet magnetized in the longitudinal geometry. 
%We show that the p-polarized incident wave leads to both p- and s-polarized reflected waves at double frequency. 
The estimations show that the suggested mechanism may explain the recent experiments \cite{Murzina_OptExpr, Kolmychek_JETP, Murzina_JETPL}, strong effect reported for a cobalt / heavy metal system may be explained by enhanced dissipation due to spin current from a ferromagnet to heavy metal \cite{Tserkovnyak_PRL, Tserkovnyak_RMP}. 
THz generation via such a magnetization dynamics is also discussed.
%Static electric field caused by the rectification effect is also found. This leads to the emission of a THz electromagnetic wave if the system is irradiated by a femtosecond laser pulse.

%%%%%%%%%%%%%%%%%%%%%%%%%%%%%%%%%%%%%%%%%%
\section{%Materials and Methods
Theoretical approach} \label{Sec_Meth}

We start with the Landau-Lifshitz-Gilbert equation for a uniform magnetic medium placed in an alternating magnetic field:
\begin{eqnarray} \label{Meth_LLG}
    \frac{d \bf{M}}{dt} &=& -\frac{\omega_M}{M_s} \left[\bf{M} \times \bf{H} + \bf{h}'\right] \\ \nonumber
    &+& \frac{\alpha \omega_M}{M_s^2} \left[\bf{M} \times \left[\bf{M} \times \bf{H} + \bf{h}'\right]\right],
\end{eqnarray}
where $\bf{M}$ is the magnetization, $M_s$~-- its saturation value, %magnetization, 
$\bf{H}$ is the external magnetic field, $\omega_M = \gamma M_s$ is the magnetization oscillation characteristic frequency, $\gamma$ is the gyromagnetic ratio, $\alpha$ is the dimensionless Gilbert damping constant of the considered medium, $\left.\bf{h}' = \bf{h}_0'\right. \cos\omega t$ is the alternating magnetic field of the optical wave %of the frequency $\omega$ %, where the stroke denotes that this field is 
inside the medium. We then %suppose 
take into account that the optical %wave 
frequency is much larger than the magnetic system eigenfrequency, $\omega >> \omega_M$. This assumption is valid as the ratio $\frc{\omega_M}{\omega}$ is typically $10^{-4}$ or less. Then the equation (\ref{Meth_LLG}) can be solved with the use of Kapitza pendulum method \cite{Kapitza}. The general approach to this problem is described below.

Let us consider a system of differential equations for arbitrary number of coordinates $A_i$ with a rapidly oscillating external source:
\begin{equation} \label{Meth_gen}
    \frac{d A_i}{dt} = f_i\left(\bf{A}\right) + g_i\left(\bf{A}\right) \cos\omega t + h_i\left(\bf{A}\right) \sin\omega t
\end{equation}
where we take into account arbitrary phase of the source in the right-hand part of equation by introducing the two sets of real functions $g_i$ and $h_i$, $\textbf{A}$ is the coordinate vector. One may determine the %several 
eigenfrequencies %that describe 
of the system (\ref{Meth_gen}) as
\begin{equation} \label{Meth_freq}
    \Omega_{f\; ij} = \frac{\partial f_i}{\partial A_j},
    \Omega_{g\; ij} = \frac{\partial g_i}{\partial A_j},
    \Omega_{h\;ij} = \frac{\partial h_i}{\partial A_j}.
\end{equation}
and suppose that they all %these frequencies 
are much smaller than that of the external source, i.e. $\Omega_{f,g,h\; ij} << \omega$. We also suppose that $\Omega_{f\; ij} \sim \Omega_{g\; ij} \sim \Omega_{h\; ij} \sim \Omega$ where $\Omega$ is the characteristic frequency of system motion. This is governed by the fact that the oscillation of the external source, which is explicitly written in (\ref{Meth_gen}) in terms of $g_i$ and $h_i$, is much faster than all the characteristic times of the system (including the time at which magnitude of the source changes). Such a supposition corresponds to a wide range of physical systems, including the system under consideration (\ref{Meth_LLG}) in which an electromagnetic wave of optical frequency acts on magnetization of a ferromagnet.

Then we may seek the solution of the system (\ref{Meth_gen}) in the form
\begin{equation} \label{Meth_sol}
    \bf{A} = \bf{U} + \bf{a}
\end{equation}
supposing that $\bf{U}$ is a ``slow'' part of the solution with the typical frequency $\Omega$ and $\bf{a}$ is  the ``fast''oscillating  part with the characteristic frequency $\omega$. It is then straightforward to split the functions $f_i, g_i, h_i$ into series:
\begin{eqnarray} \label{Meth_f}
    f_i\left(\bf{A}\right) &\approx& f_i\left(\bf{U}\right) + \sum_j{\left.\frac{\partial f_i}{\partial A_j}\right|_{\bf{U}} a_j} \\ \nonumber
    &+& \frac{1}{2} \sum_{jk}{\left.\frac{\partial^2 f_i}{\partial A_j \partial A_k}\right|_{\bf{U}} a_j a_k},
\end{eqnarray}
\begin{equation} \label{Meth_g}
    g_i\left(\bf{A}\right) \approx g_i\left(\bf{U}\right) + \sum_j{\left.\frac{\partial g_i}{\partial A_j}\right|_{\bf{U}} a_j},
\end{equation}
\begin{equation} \label{Meth_h}
    h_i\left(\bf{A}\right) \approx h_i\left(\bf{U}\right) + \sum_j{\left.\frac{\partial h_i}{\partial A_j}\right|_{\bf{U}} a_j}.
\end{equation}
Here we provide terms up to $\left(\frc{\Omega}{\omega}\right)^2$ for general solution (see below). However we need only terms linear in $\frc{\Omega}{\omega}$ to solve (\ref{Meth_LLG}) in the framework of current paper.

By substituting (\ref{Meth_f})--(\ref{Meth_h}) into (\ref{Meth_gen}) and averaging over small time period corresponding to the frequency  $\omega$ one may obtain the equation for the ``slow'' part $\bf{U}$:
\begin{eqnarray} \nonumber
    \dot{U}_i &=& f_i - \frac{1}{2 \omega} \sum_j{\left(\frac{\partial g_i}{\partial A_j} h_j - \frac{\partial h_i}{\partial A_j} g_j \right)} \\ \label{Meth_U}  
    &-& \frac{1}{2 \omega^2} \sum_{jk}{\left(\frac{\partial g_i}{\partial A_j} \frac{\partial f_i}{\partial A_k} g_k + \frac{\partial h_i}{\partial A_j} \frac{\partial f_i}{\partial A_k} h_k \right.} \\ \nonumber
    &-&\left.\frac{1}{2} \frac{\partial^2 f_i}{\partial A_j \partial A_k} \left(g_j g_k + h_j h_k\right)\right)
\end{eqnarray}
where we restrict ourselves by the second order in $\frc{\Omega}{\omega}$ and consider %suppose that 
the functions $f_i, g_i, h_i$  and their derivatives are taken at the $\bf{U}$ coordinate. Note that we suppose that second derivative of $f_i$ gives a term proportional to $\Omega^2$, e.g. $\frac{\partial^2 f_i}{\partial A_j \partial A_k} g_j \propto \Omega^2$. It is obvious from (\ref{Meth_U}) that an arbitrary shift of the phase of the oscillating source would lead to change of $g_i$ and $h_i$ while keeping $\bf{U}$ intact. The equation of motion for a classic pendulum with vibrating suspension \cite{Kapitza} is obtained from the third term of the right-hand part of (\ref{Meth_U}) ($\propto \frc{1}{\omega^2}$). % at the second order in $\frc{\Omega_{f,g,h}}{\omega}$.

Usually equation (\ref{Meth_U}) is then used to calculate the dynamics of the system averaged over ``fast'' oscillations  %the frequency 
of the source at the coordinate vector $\mathbf{U}$. This is done for the dynamics of magnetized medium in \cite{Akhiezer_FTT, Zvezdin_JETPL} and recently in \cite{Dzhezherya_JETP, Shukrinov_PRB}. However in order to find the sources of the first and second harmonics of the fast oscillating terms at $\omega$ frequency we need to consider the ``fast'' part of the solution. This can be made by using the  perturbation theory after substituting the equations (\ref{Meth_f})--(\ref{Meth_h}) into (\ref{Meth_gen}) and taking into account the Eq. (\ref{Meth_U}). % g the average values
%After substituting (\ref{Meth_f})--(\ref{Meth_h}) into (\ref{Meth_gen}) and taking into account the Eq. (\ref{Meth_U}) for the average values, one can  use the  perturbation theory in order to find the oscillating part of solution. 
The ``fast'' part oscillating at the source frequency $\omega$ is then integrated in the form:
\begin{eqnarray} \label{Meth_a}
    a_i^\omega &=& \left(\frac{g_i}{\omega} \sin \omega t - \frac{h_i}{\omega} \cos \omega t \right) \\ \nonumber
    &-& \left(\sum_j{\frac{\partial f_i}{\partial A_j} \frac{h_j}{\omega^2}} \sin \omega t  + \sum_j{\frac{\partial f_i}{\partial A_j} \frac{g_j}{\omega^2}} \cos \omega t \right),
\end{eqnarray}
where we  suppose again that the  functions $f_i, g_i, h_i$ and their derivatives are taken at the $\bf{U}$ point. It is clear from (\ref{Meth_a}) that the expression in the second bracket in the right-hand part is smaller than the first one as $\frc{\Omega}{\omega}$; we neglect all smaller terms in the solution. We may substitute the solution (\ref{Meth_a}) into series (\ref{Meth_f}), (\ref{Meth_g}), (\ref{Meth_h}) in order to obtain condition of applicability of this series expansion. Substituting the first bracket of the right-hand part of (\ref{Meth_a}) gives the term $\sim \frac{\Omega}{\omega} g_j$, $\sim \frac{\Omega}{\omega} h_j$ in the first order of the Taylor series, and the term $\sim \left(\frac{\Omega}{\omega}\right)^2 g_j$, $\sim \left(\frac{\Omega}{\omega}\right)^2 h_j$ in the second order of series, etc. Accordingly, substituting the second bracket of the right-hand part of (\ref{Meth_a}) gives a term proportional to square of the ratio of frequencies in the first order of Taylor series already. Taking for simplicity that $f_i$, $g_i$, and $h_i$ are of the same order of value, we get the expected result that the small parameter for the series expansion of (\ref{Meth_f}), (\ref{Meth_g}), (\ref{Meth_h}) is $\frc{\Omega}{\omega}$.

One may also find the ``fast'' part of the solution oscillating at the double frequency $2 \omega$. In the lowest order of the perturbation theory it takes the form:
\begin{eqnarray} \label{Meth_2a}
    a_i^{2 \omega} &=& \frac{1}{4 \omega^2} \left(\sum_j{\left(\frac{\partial h_i}{\partial A_j} h_j - \frac{\partial g_i}{\partial A_j} g_j\right)} \cos 2 \omega t \right.\\ \nonumber
    &-& \left. \sum_j{\left(\frac{\partial g_i}{\partial A_j} h_j + \frac{\partial h_i}{\partial A_j} g_j\right)} \sin 2 \omega t \right),
\end{eqnarray}
which is proportional to $\left(\frc{1}{\omega}\right)^2$. %in the lowest order.

\begin{figure}[t]
\centering{\includegraphics[width=0.8 \hsize]{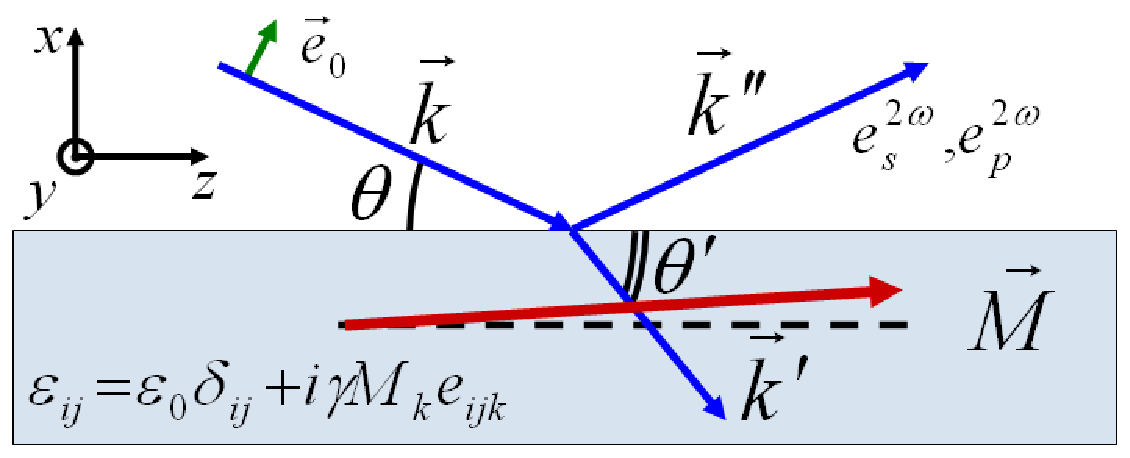}}
\caption{Geometry of the system under consideration. A p-polarized optical wave is incident at the surface of a ferromagnet magnetized in the longitudinal %initial 
geometry. Oscillations of magnetization are schematically shown as a shift of $\bf{M}$ from initial z-direction.\label{fig1}}
\end{figure}

We can now apply the general solution described above to the Landau-Lifshitz-Gilbert equation (\ref{Meth_LLG}). In order to do this, we take into account that the magnetization vector can be written through the two angles, $\varphi$ and $\beta$, with the amplitude $\left|\bf{M}\right| = M_s$, as
%and use the spherical coordinate system for $\bf{M}$:
\begin{equation} \label{Meth_M}
    \left.\bf{M}\right. = M_s \left(\cos \varphi \sin \beta, \sin \varphi \sin \beta, \cos \beta\right),
\end{equation}
in Cartesian coordinate system, as shown in Figure~\ref{fig1}.
%i.e. the magnetization vector is described by two angles, $\varphi$ and $\delta$. 
According to (\ref{Meth_a}), the  part of magnetization oscillating at the  frequency $\omega$ has a term linear in $\frc{\Omega}{\omega} \equiv \frc{\omega_M}{\omega}$. If one takes the Cartesian coordinate system in such a way that the equilibrium magnetization is parallel to z-axis and magnetic field of the wave is parallel to y-axis ($\left.\bf{h}'\right. = - e_0' \left.\bf{e}_y\right. \cos \omega t$, where $e_0'$ is the wave electric field magnitude, $\bf{e}_y$ is the unit vector in the y-direction; see Figure~\ref{fig1}) the magnetization has the following form:
\begin{eqnarray} \label{Meth_Msol}
    &\left.\bf{M}\right.& = \left.\bf{M}_0 + \bf{m}\right. = \\ \nonumber
    && \left(\frac{\omega_M}{\omega} h_0' \sin \left(\omega t - \bf{k}' \bf{r}\right), - \alpha \frac{\omega_M}{\omega} h_0' \sin \left(\omega t - \bf{k}' \bf{r}\right), M_s \right)
\end{eqnarray}
up to the first order in $\frc{\omega_M}{\omega}$. This oscillating part $\bf{m}$ of magnetization gives rise to the second harmonic generation, as we show below.

The Maxwell's equations are solved when considering the magnetization oscillations as a perturbation. We write the dielectric permittivity of the medium in the usual form:
\begin{equation} \label{Meth_epsilon}
    \varepsilon_{ij} = \varepsilon_0 \delta_{ij} + i \gamma M_k e_{ijk},
\end{equation}
where $\delta_{ij}$, $e_{ijk}$ are the Kronecker delta and the antisymmetric Levi-Civita tensor, respectively. %It is necessary 
%We have to find the real unperturbed electric field $\bf{e}'$ by solving the Maxwell's equations with $\bf{M}_0$. 
%and proper boundary conditions, and then adding the complex conjugate to the complex value. 
%This is governed by the fact that we seek for the non-linear effects in wave electric field.
The real unperturbed electric field $\bf{e}'$ is found by solving the Maxwell's equations with the magnetization $\bf{M}_0$. 
After that, the linear in $\bf{m}$ correction $\delta \bf{e}'$ %to it 
 is found as a  solution of the equation
\begin{eqnarray} \label{Meth_Maxwell}
    \nabla &\times& \left(\nabla \times \delta \left.\bf{e}'\right. \right) + \frac{\varepsilon_0}{c^2} \ddot{\delta \left.\bf{e}'\right.} \\ \nonumber
    &=& -i \frac{\gamma}{c^2} \left( \ddot{\bf{e}'} \times \bf{m} + 2 \dot{\bf{e}'} \times \dot{\bf{m}} + \bf{e}' \times \ddot{\bf{m}}\right)
\end{eqnarray}
which follows from the Maxwell's equations in the linear order in the gyrotropic component $\gamma$ of the dielectric permittivity, $c$ being the light velocity. The right-hand part of the equation (\ref{Meth_Maxwell}) acts as a source of the electric field $\delta \bf{e}'$ and is proportional to the square of unperturbed field in accordance with (\ref{Meth_Msol}), therefore it leads to the generation of the second-harmonic field. % $\delta \bf{e}'$. 
Note that equation (\ref{Meth_Maxwell}) is written for the second derivatives of $\delta \bf{e}'$, hence it does not describe the rectification effect (or the zero-frequency field). This problem is discussed below.

Let us consider the p-polarized electromagnetic wave with the electric field $\left.\bf{e} = \bf{e_0}\right. cos\left(\omega t - \bf{k r}\right)$, $\left.\bf{e}_0 = \bf{e}_x\right. \cos \theta + \left. \bf{e}_z \right. \sin \theta$  incident at the surface of a ferromagnet as shown in Figure~\ref{fig1}. Here  $\theta$ is the incident sliding angle, $\left.\bf{k}\right. = k \left(\cos \theta \left.\bf{e}_z\right. - \sin \theta \left.\bf{e}_x\right.\right)$ is the wavevector, $\bf{e}_i$ are the unit vectors of Cartesian coordinate system. There are two eigenmodes inside the medium, which have different refractive index and structure \cite{Landau8} (approximate electric field structure for the modes is written out below). Since these modes have elliptical polarization in general case both of them are excited by the p-polarized incident wave. The wavevectors of these modes inside a medium are determined from the boundary conditions  at the magnetic interface as:
\begin{equation} \label{Meth_kstr}
    k_\pm' \approx k_0 \sqrt{\varepsilon_0} \left(1 \pm \frac{\gamma M_s}{2 \varepsilon_0^{3/2}} \cos \theta\right),
\end{equation}
where $k_0 = \frac{\omega}{c}$. Corresponding sliding angles $\theta'_\pm$ inside the magnetic medium are equal to:
\begin{equation} \label{Meth_thetastr}
    \cos \theta'_\pm \approx \frac{\cos \theta}{\sqrt{\varepsilon_0}} \left(1 \mp \frac{\gamma M_s}{2 \varepsilon_0^{3/2}} \cos \theta\right).
\end{equation}
The Cartesian components of the electric field of the optical wave inside the medium may also be easily found by satisfying the boundary conditions of continuity of the tangential component of the electric field strength vector, the normal component of the electric field induction vector, and the tangential component of the magnetic field strength vector:
\begin{equation} \label{Meth_ex}
    e_{x+}' = e_{x-}' = 2 e_0 \frac{\cos \left(\omega t - \bf{k}' \bf{r}\right) \sin \theta \cos \theta}{\varepsilon_0 \sin \theta + \sqrt{\varepsilon_0 - \cos^2 \theta}},
\end{equation}
\begin{equation} \label{Meth_ey}
    e_{y \pm}' = 2 e_0 \sin \left(\omega t - \bf{k}' \bf{r}\right) \frac{\pm \sqrt{\varepsilon_0} \sin \theta + \frac{\gamma M_s}{2} \tan \theta}{\varepsilon_0 \sin \theta + \sqrt{\varepsilon_0 - \cos^2 \theta}},
\end{equation}
\begin{eqnarray} \label{Meth_ez}
    e_{z \pm}' &=& 2 e_0 \cos \left(\omega t - \bf{k}' \bf{r}\right) \\ \nonumber
    &\times& \frac{\sqrt{\varepsilon_0 - \cos^2 \theta} \sin \theta \pm \frac{\gamma M_s}{2 \sqrt{\varepsilon_0}} \frac{2 \varepsilon_0 - \cos^2 \theta}{\sqrt{\varepsilon_0 - \cos^2 \theta}} \tan \theta}{\varepsilon_0 \sin \theta + \sqrt{\varepsilon_0 - \cos^2 \theta}}.
\end{eqnarray}
Note that this solution is an approximate one and is applicable only when the terms proportional to $\gamma M_s$ are small. %(i.e. does not work for $\theta \to \frc{\pi}{2}$ when $\tan \theta \to \infty$). 
Solution (\ref{Meth_Msol}) contains the magnitude of the unperturbed  magnetic field wave inside the medium, which is determined from the boundary conditions as:
\begin{equation} \label{Meth_hstr}
    h_0' = e_0 \frac{4 \varepsilon_0 \sin \theta}{\varepsilon_0 \sin \theta + \sqrt{\varepsilon_0 -\cos^2 \theta}}.
\end{equation}

Substituting (\ref{Meth_ex})--(\ref{Meth_ez}) and (\ref{Meth_Msol}) into eq. (\ref{Meth_Maxwell}), we then solve this equation and finally find the double-frequency electric field outside the medium from the boundary conditions. The s- and p-component of its magnitude have the form:
\begin{equation} \label{Meth_res_s}
    e_s^{2 \omega} = \frac{e_0^2}{M_s} \frac{\omega_M}{\omega} \frac{\gamma M_s}{2} \frac{\varepsilon_0 \sqrt{\varepsilon_0 - \cos^2 \theta} \tan^2 \theta}{\left(\varepsilon_0 \sin \theta + \sqrt{\varepsilon_0 - \cos^2 \theta}\right)^2},
\end{equation}
\begin{equation} \label{Meth_res_p}
    e_p^{2 \omega} = \alpha \frac{e_0^2}{M_s} \frac{\omega_M}{\omega} \frac{\gamma M_s}{2} \frac{\left(\varepsilon_0 - \cos^2 \theta\right) \sin \theta \cos \theta}{\varepsilon_0 \left(\varepsilon_0 \sin \theta + \sqrt{\varepsilon_0 - \cos^2 \theta}\right)^2}.
\end{equation}
Second harmonic field determined by its p- and s-components, (\ref{Meth_res_s}) and (\ref{Meth_res_p}), appears due to oscillations of magnetization of the ferromagnet in the magnetic field of the light wave. This is the main result of current paper; it is discussed in Section~\ref{Sec_Res}.

As we have mentioned above, the static (zero-frequency) electric field is not described by eq.~(\ref{Meth_Maxwell}). However second-order nonlinear optical effects such as SHG and rectification typically coexist. In order to show that the rectified signal appears in our case as well, we provide a simple model in which the electron motion is described by the Newton's law. This method is very similar to one used by Gaponov and Miller in order to calculate the ponderomotive force that acts on a charged particle in an electromagnetic field of high frequency \cite{Gaponov_JETP}. We suppose that ``free'' conduction electrons are in charge of the optical response of the system. Their motion is described as:
\begin{equation} \label{Meth_Newton}
    \ddot{\left.\bf{r}\right.} = -\frac{e}{m_e} \left.\bf{e}'\right. - \frac{\lambda}{m_e} \dot{\bf{r}} \times \bf{M}\left(t\right),
\end{equation}
where $\bf{r}$ is the electron coordinate, $e$ is its absolute charge, $m_e$~is its mass and $\lambda$ is the constant of Lorentz-like force induced by the magnetization, which leads e.g. to anomalous Hall effect and has spin-orbit roots \cite{Sinova_RMP}. Supposing that $\lambda$ is small, we first solve the equation (\ref{Meth_Newton}) neglecting the Lorentz-like force. At the next step we substitute the obtained  solution into this force in order and find the corresponding correction to $\bf{r}\left(t\right)$. Averaging this force over time period of wave with (\ref{Meth_Msol}) gives an effective electric field that acts on the electrons:
\begin{equation} \label{Meth_Eeff}
    \left.\bf{E}_{eff}\right. = -\gamma \frac{\omega}{\omega_p^2} \left<\dot{\bf{e}'} \times \bf{m}\right>_t,
\end{equation}
where the Lorentz-like force constant $\lambda$ is expressed through the medium constant of gyrotropy $\gamma$, $\omega_p$ is the electron plasma frequency, and $\left<...\right>_t$ stands for averaging over time. After averaging with $\bf{m}$ determined by (\ref{Meth_Msol}) and $\bf{e}'$ defined as (\ref{Meth_ex})--(\ref{Meth_ez}) we obtain:
\begin{eqnarray} \label{Meth_Eeff_sol_xy}
    E_{eff\,x} &=& -\alpha E_{eff\,y} \\ \nonumber 
    &=& -\alpha \frac{e_0^2}{M_s} \frac{\omega \omega_M}{\omega_p^2} \gamma M_s \frac{4 \varepsilon_0 \sin^2 \theta \sqrt{\varepsilon_0 - \cos^2 \theta}}{\left(\varepsilon_0 \sin \theta + \sqrt{\varepsilon_0 - \cos^2 \theta}\right)^2},
\end{eqnarray}
\begin{equation} \label{Meth_Eeff_sol_z}
    E_{eff\,z} = -\alpha \frac{e_0^2}{M_s} \frac{\omega \omega_M}{\omega_p^2} \gamma M_s \frac{4 \varepsilon_0 \sin^2 \theta \cos \theta}{\left(\varepsilon_0 \sin \theta + \sqrt{\varepsilon_0 - \cos^2 \theta}\right)^2}.
\end{equation}
Thus we have the rectification effect due to magnetization oscillations inside a medium. This is the second main result of the current paper.

%%%%%%%%%%%%%%%%%%%%%%%%%%%%%%%%%%%%%%%%%%
\section{Results and Discussion} \label{Sec_Res}
The main results of our consideration presented in Section~\ref{Sec_Meth} are the equations for the s- and p-polarized components of the electric field of the SHG wave  (\ref{Meth_res_s}), (\ref{Meth_res_p}) and the rectified field (\ref{Meth_Eeff_sol_xy}), (\ref{Meth_Eeff_sol_z}) driven by oscillations of magnetic moment in the medium under the influence of p-polarized incident wave in the longitudinal geometry. We analyze these equations below.

\subsection{Second harmonic generation}
It is known \cite{Rzhevsky_PRB} that for the p-polarized light incident at the surface of an isotropic ferromagnet there is only  p-polarized non-magnetic SHG response, while only transversal component of magnetization may give rise to the p-polarized magnetic SHG signal (see Table~\ref{Table}). There is only s-polarized magnetization-induced SHG for both polar and longitudinal geometries of the experiment. This is governed by the symmetry of the surface of a ferromagnet at which the inversion symmetry is broken, as the SHG polarization $\bf{P}^{2 \omega}$ can be fully described by the following expression:
\begin{eqnarray} \label{Res_P}
    \left.\bf{P}\right.^{2\omega} &\propto& \bf{n} \bf{e}^2 + \bf{e} \left(\bf{n} \cdot \bf{e}\right) \\ \nonumber 
    &+& \bf{n} \times \bf{e} \, \left(\bf{M} \cdot \bf{e}\right) + \bf{M} \times \bf{e} \, \left(\bf{n} \cdot \bf{e}\right) + \bf{n} \times \bf{M} \, \bf{e}^2,
\end{eqnarray}
where  %is the double-frequency polarization of the medium, 
$\bf{n}$ is the surface normal vector and $\bf{e}$~is the magnitude of the electric field of the incident wave.% as above.
The first two terms in the right-hand part of (\ref{Res_P}) stand for the nonmagnetic response, while the last three ones are linear in $\bf{M}$. It is then straightforward to obtain the results summarized in Table~\ref{Table}. One can see that the p-polarized SHG is ``forbidden'' %by symmetry 
for $M_y = 0$. %, i.e. there is no magnetization perpendicular to the plane of incidence of the wave.

\begin{table}[bt] 
\caption{Contributions to SHG appearing in different combinations
of the polarizations of the exciting and SHG light. ``+'' or ``-'' represent existence or absence of SHG light, respectively. SHG discovered in current paper is marked after slash where applicable.\label{Table}}
%\newcolumntype{C}{>{\centering\arraybackslash}X}
\begin{tabular}{c|c|c|c|c}
Polarization& $M_x$	& $M_y$ & $M_z$ & nonmagnetic\\
\toprule
$P_{in} P_{out}$		& -			& +     & -/+     & +     \\
$P_{in} S_{out}$		& +			& -     & +/+     & -     \\ 
$S_{in} P_{out}$		& -			& +     & -     & +     \\ 
$S_{in} S_{out}$		& -			& -     & +     & -     
\end{tabular}
%\noindent{\footnotesize{\textsuperscript{1} Tables may have a footer.}}
\end{table}

The second harmonic generation (\ref{Meth_res_s}),(\ref{Meth_res_p}) discovered in this paper is of different nature. It  is governed by the oscillation of magnetization of the magnetic field $\bf{h}'$ inside the medium together with the nonlinearity of the material equation for the electric induction; it contains the vector product of the electric field $\bf{e}'$ and the oscillating magnetization. This mechanism is not related to the break of the inversion symmetry at the interface, instead it utilizes the break of this symmetry by the wavevector $\bf{k}' \propto \bf{e}' \times \bf{h}'$. Therefore it removes the symmetry  restriction on magnetization-induced effect in SHG for the $P_{in} P_{out}$ combination of polarizations illustrated by Table~\ref{Table}. We suppose that the same would apply to $S_{in} P_{out}$ polarization combination.

\begin{figure}[t]
\centering{\includegraphics[width=0.9 \hsize]{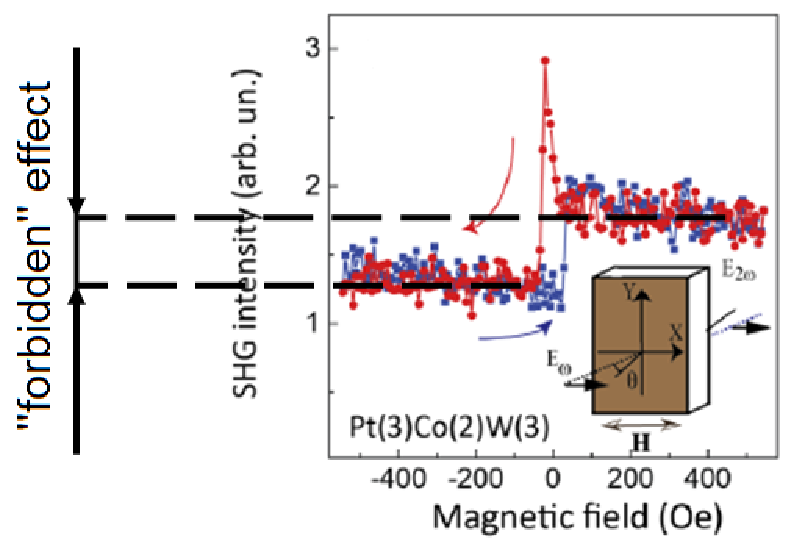}}
\caption{Dependence of the SHG intensity for the $P_{in} P_{out}$ polarizaton combination on longitudinal magnetic field for the sliding angle $\theta = 70^\circ$ for Pt(3nm)/Co(3nm)/W(3nm) film. The picture is taken from \cite{Murzina_OptExpr}. (The Cartesian coordinate x here corresponds to the coordinate z in current paper.) \label{fig2}}
\end{figure}

The discovered ``forbidden'' effect was earlier observed in \cite{Murzina_OptExpr}; the main result of this experiment is shown in Figure~\ref{fig2}. 
Here the SHG intensity hysteresis in the longitudinal geometry measured  for the $P_{in} P_{out}$ polarizations' combination shows a clear difference in the SHG signal for the positive and negative saturating magnetic field. %Properties in small magnetic field may be attributed to process of re-magnetization through some intermediate non-uniform state. 
This ``forbidden'' effect %shown in the picture 
may be explained by the mechanism discussed in the current paper. 

This is supported by the following estimations. In the experiment, the pulsed laser radiation at $820 nm$ wavelength  with $30 fs$ pulse duration is used. The peak pulse power is about $70kW$, the beam diameter is $30 \mu m$,  which gives the intensity of $\sim 10^{10} \frac{W}{cm^2}$ and the electric field of $2.7 \cdot 10^6 \frac{V}{cm}$. As the saturation magnetization of cobalt is $1400G$, we get the ratio  %and therefore the ratio of the wave magnetic field to $M_s$ is 
$\frc{h_0}{M_s} \approx 6.5$. We can estimate the gyrotropic term of the dielectric permittivity from the MOKE polarization rotation angle, which is  about $5 \cdot 10^{-3} rad$ for Co films. By taking the approximate value $\varepsilon_0 \sim 10$ by the order of value, we obtain $\gamma M_s \approx 0.05$, which is a small parameter indeed. 

The frequency ratio which is the main small parameter that determines the magnetization oscillation magnitude is $\frc{\omega_M}{\omega} \approx 0.6 \cdot 10^{-4}$, and the Gilbert damping constant is approximately $\alpha \sim 0.1$ by the order of value for a Co~/~Pt system (see below). Using eq. (\ref{Meth_res_p}) and neglecting the angular dependence, we arrive at the estimation $e_p^{2 \omega} \approx 10^{-8} e_0$. Typical SHG efficiency for a ferromagnetic surface is $I^{2 \omega}_s \approx 10^{-14} I^\omega$, hence the electric field $e^{2 \omega} \approx 10^{-7} e_0$ \cite{Aktsipetrov_JOSAB}. Thus the interference of the discovered magnetic SHG with the non-magnetic SHG response  from the surface gives $I^{2 \omega} \approx 10^{-15} I^\omega$, which is only an order of magnitude smaller than that for the non-magnetic signal: $\frc{I^{2 \omega}}{I^{2 \omega}_s} \approx 0.1$. 
The experimentally observed ``forbidden'' effect that can be compared to our estimations is determined as $\frac{I^{2 \omega}\left(+H\right) - I^{2 \omega}\left(-H\right)}{I^{2 \omega}\left(+H\right) + I^{2 \omega}\left(-H\right)}$ and is approximately $0.17$. Thus the suggested mechanism gives the value of the same order of magnitude as observed in the experiment.

According to \cite{Murzina_OptExpr}, the observed ``forbidden'' magnetization-induced SHG intensity effect decreases as the Co layer thickness grows in a Co/Pt or Co/W bilayer film. %and is smaller than noise for $10nm$. 
This is consistent with the fact that the discussed SHG effect is proportional to the Gilbert damping constant $\alpha$, which is enhanced in an FM/HM system due to the spin current flow at the ferromagnet/heavy metal  interface \cite{Tserkovnyak_PRL}. As this is a surface effect, it decreases as the cobalt thickness grows. Accordingly, the ``forbidden'' SHG effect discovered in this paper decreases. 

Table~\ref{Table2} summarizes the results of  rough estimations of the ``forbidden'' SHG effect for different materials. One can see that the increase of saturation magnetization and of the Gilbert damping constant leads to the increase of the effect. So the best choice for its observation is a thin Co~/~Pt multilayer system, which stays in agreement with the mentioned experiments.
\begin{table}[hbt] 
\caption{Typical parameters for different ferromagnetic materials and estimations for the ``forbidden'' $P_{in} P_{out}$ SHG effect.\label{Table2}}
%\newcolumntype{C}{>{\centering\arraybackslash}X}
\begin{tabular}{c|c|c|c|c|c}
Material& $M_s, G$	& $\gamma M_s$ & $\alpha$ & $\frc{I^{2 \omega}}{I^{2 \omega}_s}$ & Ref.\\
\toprule
YIG		& 200			& $5 \cdot 10^{-4}$   & $2.3 \cdot 10^{-4}$ & $2.3 \cdot 10^{-6}$     & \cite{Yu_SciRep}     \\
$Ni_{80}Fe_{20}$		& 800			& $0.02$   & $0.01$ & $4 \cdot 10^{-3}$     & \cite{Zhang_AIP,Dorofey_JTP}     \\ 
CoFeB		& 1200			& $0.04$    & $0.015$ & $1.2 \cdot 10^{-2}$     & \cite{Dorofey_JTP}     \\ 
Co		& 1400			& $0.05$   & $0.02$ & $2 \cdot 10^{-2}$     & \cite{Skorokhodov_JETP}     \\ 
thin~Co/Pt		& 1400		& $0.05$    & $0.04-0.22$ & $0.04-0.22$     & \cite{Tserkovnyak_PRL}  
\end{tabular}
%\noindent{\footnotesize{\textsuperscript{1} Tables may have a footer.}}
\end{table}

\begin{figure}[hbt]
\centering{\includegraphics[width=0.9 \hsize]{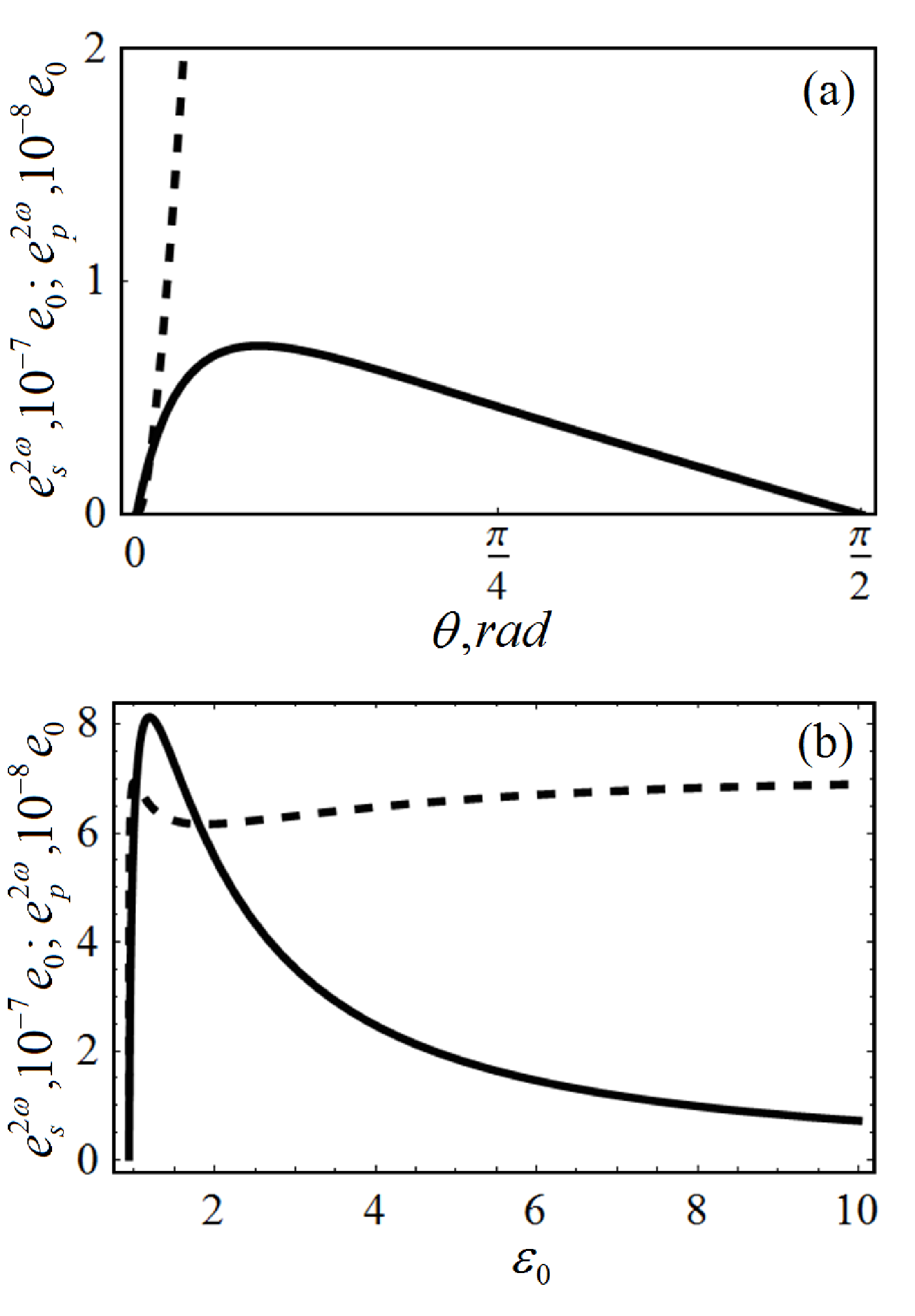}}
\caption{Dependence of electric field of the p-polarized (solid line) and s-polarized (dashed line) second-harmonic wave on (a) sliding angle for $\varepsilon_0 = 10$ and (b) dielectric permittivity for $\theta = 15^\circ$. Other parameters are the same as mentioned in the text for thin Co~/~Pt system. \label{fig3}}
\end{figure}

The dependences of the calculated SHG fields (\ref{Meth_res_s}), (\ref{Meth_res_p}) on the sliding angle $\theta$ and the dielectric permittivity $\varepsilon_0$ of the medium are shown in Figure~\ref{fig3}. The field of the s-polarized SHG wave grows as the sliding angle increases. It diverges at $\theta = \frac{\pi}{2}$ when % which is explained by the fact that
the %approximate 
solution (\ref{Meth_res_s}) is incorrect for $\tan \theta \to \infty$. The electric field of the p-polarized SHG wave reveals a maximum at $\theta \approx 15^\circ$ sliding angle. This field has a maximum at quite small dielectric permittivity ($\varepsilon_0 \approx 1.2$) and, contrary to the field of the s-polarized SHG wave, decreases as $\frc{1}{\varepsilon_0}$ for $\varepsilon_0 >> 1$.

%\subsubsection{Subsubsection}

%Bulleted lists look like this:
%\begin{itemize}
%\item	First bullet;
%\item	Second bullet;
%\item	Third bullet.
%\end{itemize}

%Numbered lists can be added as follows:
%\begin{enumerate}
%\item	First item; 
%\item	Second item;
%\item	Third item.
%\end{enumerate}

\subsection{Rectification effect}
The effective rectified field that appears due to magnetization oscillations under the subjection of the magnetic field of the optical wave is determined by (\ref{Meth_Eeff_sol_xy}), (\ref{Meth_Eeff_sol_z}). We may estimate this field for realistic parameters of a femtosecond laser pulse described above. The plasma frequency of a metal with the electron concentration of  $10^{22} cm^{-3}$ is $\omega_p = 5.6 \cdot 10^{15} s^{-1}$. The angle function in (\ref{Meth_Eeff_sol_xy}) reaches its maximum at $\theta \to \frc{\pi}{2}$, which corresponds to normal incidence of the light wave. Substituting all parameters into (\ref{Meth_Eeff_sol_xy}), (\ref{Meth_Eeff_sol_z}) we obtain %the estimation 
$E_{eff \, y} \approx 10^{-6} e_0 \approx 3.7 \frac{V}{cm}$. This effect is relatively weak, while this value can be increased by lowering  the plasma frequency, e.g. by taking diluted magnetic semiconductors.

We note that z-component of the effective field is caused by the x-component of the light wave. Therefore it should be zero due to screening effects that are not taken into account in (\ref{Meth_Eeff_sol_xy}), (\ref{Meth_Eeff_sol_z}). On the other hand, the x-component of the effective field would be cancelled due to the same screening effect. Therefore the effective field has only y-component, which is larger than the other components since it does not contain the small damping factor $\alpha$. Taking this into account, from a symmetry point of view the effective field may be written as 
\begin{equation} \label{Res_Eeff}
    \bf{E}_{eff} \sim \bf{M} \times \bf{n}.
\end{equation}

The electric current caused by this effective field may be determined from  the Ohm's law $\left.\bf{j}\right. = \sigma \bf{E}_{eff}$. For the  thickness of $30$ nm and the width of the  the current flow area  equal to the beam diameter  of 30 $\mu m$ we estimate the constant electric current as $I_e \approx 4 mA$. For $80 MHz$ pulse repetition rate this gives the average current $\left<I_e\right> \approx 1 nA$, which can hardly be detected in real systems. However this current appears at an electromagnetic wave envelope time which is usually $30-50 ps$ and thus should emit the THz radiation with the characteristic frequency of $20-33$ THz. In real systems, characteristic time of the electric current relaxation is determined by the electron-phonon interaction and is of the order of 100 fs - 1 ps. Thus the electromagnetic wave generation is usually restricted to several THz. The polarization of a THz wave is determined by (\ref{Res_Eeff}) and is the same as conventional for the spintronic THz emitters \cite{Bull_APLM}. However this additional mechanism does not depend on the constant of Gilbert damping and therefore should exist as well for a single ferromagnetic layer. Contrary to the mechanism that provides THz generation in ferromagnetic/nonmagnetic systems, the effect discovered here should give the signal growing with the thickness of the FM layer.

\section{Conclusions}

In conclusion, we theoretically investigate the non-linear optical effects that appear due to magnetization oscillations under the influence of the %magnetic field of the 
optical wave %incident at 
on a ferromagnet surface. Based on the Kapitza pendulum approach  we show that the light-induced magnetization dynamics in a ferromagnet can provide a mechanism for the second harmonic generation.  Although the magnitude of the magnetization oscillations is small, laser-induced magnetization dynamics can provide the SHG response comparable to the nonmagnetic one, as well as to the THz generation through the rectification effect. This suggested SHG mechanism can explain the recently observed  magnetooptical SHG effect in the geometry of the longitudinal magnetooptical effect and for the $P_{in} P_{out}$ combination  of polarizations of the incident and scattered SHG waves, which is symmetry forbidden in a ferromagnetic medium. Such a ``forbidden'' effect is a consequence of damping of magnetization oscillations in a magnetic system. 
%We also show that the nonlinear-optical rectification  effect can lead to the terahertz waves generation in a ferromagnet irradiated by femtosecond laser pulses.

%The rectification effect and second harmonic generation are considered. 
%Although the magnitude of the magnetization oscillations is rather small, the laser-induced magnetization dynamics can provide the SHG response comparable to the nonmagnetic one. The nonlinear-optical rectification %effect 
%provides a mechanism for the terahertz waves generation in a ferromagnet irradiated by femtosecond laser pulses. The discussed SHG mechanism can be responsible for the SHG magnetic field induced effect for the $P_{in} P_{out}$ combination  of polarizations of the incident and scattered SHG waves, which is symmetry forbidden in  ferromagnetic media.
%if the inversion symmetry break is attributed to the surface of the ferromagnet. 
%Such a ``forbidden'' effect is a consequence of damping of magnetization oscillations in the system. 
%It was recently observed in an expreiment in a Pt(3nm)/Co(3nm)/W(3nm) multilayer system. 
%Our estimations show that the mechanism investigated in current paper may explain the results of these experiments. Strong decrease of this effect as the thickness of Co layer grows to $10nm$ is explained by enhancement of Gilbert damping in a ferromagnet~/~heavy metal system which is more pronounced for thin ferromagnetic layer.

\begin{acknowledgments}
This work is supported by the Russian Science Foundation, grant No. 23-22-00295.
\end{acknowledgments}

\nocite{*}

\bibliography{SHG_magn_field}% Produces the bibliography via BibTeX.

\providecommand{\noopsort}[1]{}\providecommand{\singleletter}[1]{#1}%
\begin{thebibliography}{27}%
\makeatletter
\providecommand \@ifxundefined [1]{%
 \@ifx{#1\undefined}
}%
\providecommand \@ifnum [1]{%
 \ifnum #1\expandafter \@firstoftwo
 \else \expandafter \@secondoftwo
 \fi
}%
\providecommand \@ifx [1]{%
 \ifx #1\expandafter \@firstoftwo
 \else \expandafter \@secondoftwo
 \fi
}%
\providecommand \natexlab [1]{#1}%
\providecommand \enquote  [1]{``#1''}%
\providecommand \bibnamefont  [1]{#1}%
\providecommand \bibfnamefont [1]{#1}%
\providecommand \citenamefont [1]{#1}%
\providecommand \href@noop [0]{\@secondoftwo}%
\providecommand \href [0]{\begingroup \@sanitize@url \@href}%
\providecommand \@href[1]{\@@startlink{#1}\@@href}%
\providecommand \@@href[1]{\endgroup#1\@@endlink}%
\providecommand \@sanitize@url [0]{\catcode `\\12\catcode `\$12\catcode `\&12\catcode `\#12\catcode `\^12\catcode `\_12\catcode `\%12\relax}%
\providecommand \@@startlink[1]{}%
\providecommand \@@endlink[0]{}%
\providecommand \url  [0]{\begingroup\@sanitize@url \@url }%
\providecommand \@url [1]{\endgroup\@href {#1}{\urlprefix }}%
\providecommand \urlprefix  [0]{URL }%
\providecommand \Eprint [0]{\href }%
\providecommand \doibase [0]{https://doi.org/}%
\providecommand \selectlanguage [0]{\@gobble}%
\providecommand \bibinfo  [0]{\@secondoftwo}%
\providecommand \bibfield  [0]{\@secondoftwo}%
\providecommand \translation [1]{[#1]}%
\providecommand \BibitemOpen [0]{}%
\providecommand \bibitemStop [0]{}%
\providecommand \bibitemNoStop [0]{.\EOS\space}%
\providecommand \EOS [0]{\spacefactor3000\relax}%
\providecommand \BibitemShut  [1]{\csname bibitem#1\endcsname}%
\let\auto@bib@innerbib\@empty
%</preamble>
\bibitem [{\citenamefont {Shen}(1984)}]{Shen}%
  \BibitemOpen
  \bibfield  {author} {\bibinfo {author} {\bibfnamefont {Y.~R.}\ \bibnamefont {Shen}},\ }\href@noop {} {\emph {\bibinfo {title} {The Principles of Nonlinear Optics}}}\ (\bibinfo  {publisher} {Wiley},\ \bibinfo {year} {1984})\BibitemShut {NoStop}%
\bibitem [{\citenamefont {Aktsipetrov}(2011)}]{Aktsipetrov_JOSAB}%
  \BibitemOpen
  \bibfield  {author} {\bibinfo {author} {\bibfnamefont {O.~A.}\ \bibnamefont {Aktsipetrov}},\ }\bibfield  {title} {\bibinfo {title} {Surface nonlinear optics and nonlinear magneto- optics at moscow state university},\ }\href@noop {} {\bibfield  {journal} {\bibinfo  {journal} {J. Opt. Soc. Am. B}\ }\textbf {\bibinfo {volume} {28}},\ \bibinfo {pages} {A27} (\bibinfo {year} {2011})}\BibitemShut {NoStop}%
\bibitem [{\citenamefont {Kim}\ \emph {et~al.}(2008)\citenamefont {Kim}, \citenamefont {Taylor}, \citenamefont {Glownia},\ and\ \citenamefont {Rodrigues}}]{Kim_NatPhot}%
  \BibitemOpen
  \bibfield  {author} {\bibinfo {author} {\bibfnamefont {K.~Y.}\ \bibnamefont {Kim}}, \bibinfo {author} {\bibfnamefont {A.~J.}\ \bibnamefont {Taylor}}, \bibinfo {author} {\bibfnamefont {J.~H.}\ \bibnamefont {Glownia}},\ and\ \bibinfo {author} {\bibfnamefont {G.}~\bibnamefont {Rodrigues}},\ }\bibfield  {title} {\bibinfo {title} {Coherent control of terahertz supercontinuum generation in ultrafast laser–gas interactions},\ }\href@noop {} {\bibfield  {journal} {\bibinfo  {journal} {Nat. Photonics}\ }\textbf {\bibinfo {volume} {2}},\ \bibinfo {pages} {605} (\bibinfo {year} {2008})}\BibitemShut {NoStop}%
\bibitem [{\citenamefont {Gildenburg}\ and\ \citenamefont {Vvedenskii}(2007)}]{Gildenburg_PRL}%
  \BibitemOpen
  \bibfield  {author} {\bibinfo {author} {\bibfnamefont {V.~B.}\ \bibnamefont {Gildenburg}}\ and\ \bibinfo {author} {\bibfnamefont {N.~V.}\ \bibnamefont {Vvedenskii}},\ }\bibfield  {title} {\bibinfo {title} {Optical-to-thz wave conversion via excitation of plasma oscillations in the tunneling-ionization process},\ }\href@noop {} {\bibfield  {journal} {\bibinfo  {journal} {Phys.\ Rev.\ Lett.}\ }\textbf {\bibinfo {volume} {98}},\ \bibinfo {pages} {245002} (\bibinfo {year} {2007})}\BibitemShut {NoStop}%
\bibitem [{\citenamefont {Vvedenskii}\ \emph {et~al.}(2014)\citenamefont {Vvedenskii}, \citenamefont {A.~I.~Korytin}, \citenamefont {Kostin}, \citenamefont {Murzanev}, \citenamefont {Silaev},\ and\ \citenamefont {Stepanov}}]{Vvedenskii_PRL}%
  \BibitemOpen
  \bibfield  {author} {\bibinfo {author} {\bibfnamefont {N.~V.}\ \bibnamefont {Vvedenskii}}, \bibinfo {author} {\bibfnamefont {A.}~\bibnamefont {A.~I.~Korytin}}, \bibinfo {author} {\bibfnamefont {V.~A.}\ \bibnamefont {Kostin}}, \bibinfo {author} {\bibfnamefont {A.~A.}\ \bibnamefont {Murzanev}}, \bibinfo {author} {\bibfnamefont {A.~A.}\ \bibnamefont {Silaev}},\ and\ \bibinfo {author} {\bibfnamefont {A.~N.}\ \bibnamefont {Stepanov}},\ }\bibfield  {title} {\bibinfo {title} {Optical-to-thz wave conversion via two-color laser-plasma generation of terahertz radiation using a frequency-tunable half harmonic of a femtosecond pulse},\ }\href@noop {} {\bibfield  {journal} {\bibinfo  {journal} {Phys.\ Rev.\ Lett.}\ }\textbf {\bibinfo {volume} {112}},\ \bibinfo {pages} {055004} (\bibinfo {year} {2014})}\BibitemShut {NoStop}%
\bibitem [{\citenamefont {Kampfrath}\ \emph {et~al.}(2013)\citenamefont {Kampfrath}, \citenamefont {Battiato}, \citenamefont {Maldonado}, \citenamefont {Eilers}, \citenamefont {N{\"o}tzold}, \citenamefont {M{\"a}hrlein}, \citenamefont {Zbarsky}, \citenamefont {Freimuth}, \citenamefont {Mokrousov}, \citenamefont {Bl{\"u}gel}, \citenamefont {Wolf}, \citenamefont {Radu}, \citenamefont {Oppeneer},\ and\ \citenamefont {M{\"u}nzenberg}}]{Kampfrath_NatNano}%
  \BibitemOpen
  \bibfield  {author} {\bibinfo {author} {\bibfnamefont {T.}~\bibnamefont {Kampfrath}}, \bibinfo {author} {\bibfnamefont {M.}~\bibnamefont {Battiato}}, \bibinfo {author} {\bibfnamefont {P.}~\bibnamefont {Maldonado}}, \bibinfo {author} {\bibfnamefont {G.}~\bibnamefont {Eilers}}, \bibinfo {author} {\bibfnamefont {J.}~\bibnamefont {N{\"o}tzold}}, \bibinfo {author} {\bibfnamefont {S.}~\bibnamefont {M{\"a}hrlein}}, \bibinfo {author} {\bibfnamefont {V.}~\bibnamefont {Zbarsky}}, \bibinfo {author} {\bibfnamefont {F.}~\bibnamefont {Freimuth}}, \bibinfo {author} {\bibfnamefont {Y.}~\bibnamefont {Mokrousov}}, \bibinfo {author} {\bibfnamefont {S.}~\bibnamefont {Bl{\"u}gel}}, \bibinfo {author} {\bibfnamefont {M.}~\bibnamefont {Wolf}}, \bibinfo {author} {\bibfnamefont {I.}~\bibnamefont {Radu}}, \bibinfo {author} {\bibfnamefont {P.~M.}\ \bibnamefont {Oppeneer}},\ and\ \bibinfo {author} {\bibfnamefont {M.}~\bibnamefont {M{\"u}nzenberg}},\ }\bibfield  {title} {\bibinfo {title} {Terahertz spin current pulses controlled by
  magnetic heterostructures},\ }\href@noop {} {\bibfield  {journal} {\bibinfo  {journal} {Nat. Nanotechnol.}\ }\textbf {\bibinfo {volume} {8}},\ \bibinfo {pages} {256} (\bibinfo {year} {2013})}\BibitemShut {NoStop}%
\bibitem [{\citenamefont {Seifert}\ \emph {et~al.}(2016)\citenamefont {Seifert}, \citenamefont {Jaiswal}, \citenamefont {Martens}, \citenamefont {Hannegan}, \citenamefont {Braun}, \citenamefont {Maldonado}, \citenamefont {Freimuth}, \citenamefont {Kronenberg}, \citenamefont {Henrizi}, \citenamefont {Radu}, \citenamefont {Beaurepaire}, \citenamefont {Mokrousov}, \citenamefont {Oppeneer}, \citenamefont {Jourdan}, \citenamefont {Jakob}, \citenamefont {Turchinovich}, \citenamefont {Hayden}, \citenamefont {Wolf}, \citenamefont {M{\"u}nzenberg}, \citenamefont {Kl{\"a}ui},\ and\ \citenamefont {Kampfrath}}]{Seifert_NatPhot}%
  \BibitemOpen
  \bibfield  {author} {\bibinfo {author} {\bibfnamefont {T.}~\bibnamefont {Seifert}}, \bibinfo {author} {\bibfnamefont {S.}~\bibnamefont {Jaiswal}}, \bibinfo {author} {\bibfnamefont {U.}~\bibnamefont {Martens}}, \bibinfo {author} {\bibfnamefont {J.}~\bibnamefont {Hannegan}}, \bibinfo {author} {\bibfnamefont {L.}~\bibnamefont {Braun}}, \bibinfo {author} {\bibfnamefont {P.}~\bibnamefont {Maldonado}}, \bibinfo {author} {\bibfnamefont {F.}~\bibnamefont {Freimuth}}, \bibinfo {author} {\bibfnamefont {A.}~\bibnamefont {Kronenberg}}, \bibinfo {author} {\bibfnamefont {J.}~\bibnamefont {Henrizi}}, \bibinfo {author} {\bibfnamefont {I.}~\bibnamefont {Radu}}, \bibinfo {author} {\bibfnamefont {E.}~\bibnamefont {Beaurepaire}}, \bibinfo {author} {\bibfnamefont {Y.}~\bibnamefont {Mokrousov}}, \bibinfo {author} {\bibfnamefont {P.~M.}\ \bibnamefont {Oppeneer}}, \bibinfo {author} {\bibfnamefont {M.}~\bibnamefont {Jourdan}}, \bibinfo {author} {\bibfnamefont {G.}~\bibnamefont {Jakob}}, \bibinfo {author} {\bibfnamefont
  {D.}~\bibnamefont {Turchinovich}}, \bibinfo {author} {\bibfnamefont {L.~M.}\ \bibnamefont {Hayden}}, \bibinfo {author} {\bibfnamefont {M.}~\bibnamefont {Wolf}}, \bibinfo {author} {\bibfnamefont {M.}~\bibnamefont {M{\"u}nzenberg}}, \bibinfo {author} {\bibfnamefont {M.}~\bibnamefont {Kl{\"a}ui}},\ and\ \bibinfo {author} {\bibfnamefont {T.}~\bibnamefont {Kampfrath}},\ }\bibfield  {title} {\bibinfo {title} {Efficient metallic spintronic emitters of ultrabroadband terahertz radiation},\ }\href@noop {} {\bibfield  {journal} {\bibinfo  {journal} {Nat. Photonics}\ }\textbf {\bibinfo {volume} {10}},\ \bibinfo {pages} {483} (\bibinfo {year} {2016})}\BibitemShut {NoStop}%
\bibitem [{\citenamefont {Bull}\ \emph {et~al.}(2021)\citenamefont {Bull}, \citenamefont {Hewett}, \citenamefont {Ji}, \citenamefont {Lin}, \citenamefont {Thomson}, \citenamefont {Graham},\ and\ \citenamefont {Nutter}}]{Bull_APLM}%
  \BibitemOpen
  \bibfield  {author} {\bibinfo {author} {\bibfnamefont {C.}~\bibnamefont {Bull}}, \bibinfo {author} {\bibfnamefont {S.~M.}\ \bibnamefont {Hewett}}, \bibinfo {author} {\bibfnamefont {R.}~\bibnamefont {Ji}}, \bibinfo {author} {\bibfnamefont {C.-H.}\ \bibnamefont {Lin}}, \bibinfo {author} {\bibfnamefont {T.}~\bibnamefont {Thomson}}, \bibinfo {author} {\bibfnamefont {D.~M.}\ \bibnamefont {Graham}},\ and\ \bibinfo {author} {\bibfnamefont {P.~W.}\ \bibnamefont {Nutter}},\ }\bibfield  {title} {\bibinfo {title} {Spintronic terahertz emitters: Status and prospects from a materials perspective},\ }\href@noop {} {\bibfield  {journal} {\bibinfo  {journal} {APL Mater.}\ }\textbf {\bibinfo {volume} {9}},\ \bibinfo {pages} {090701} (\bibinfo {year} {2021})}\BibitemShut {NoStop}%
\bibitem [{\citenamefont {Pan}\ \emph {et~al.}(1989)\citenamefont {Pan}, \citenamefont {Wei},\ and\ \citenamefont {Shen}}]{Pan_PRB}%
  \BibitemOpen
  \bibfield  {author} {\bibinfo {author} {\bibfnamefont {R.-P.}\ \bibnamefont {Pan}}, \bibinfo {author} {\bibfnamefont {H.~D.}\ \bibnamefont {Wei}},\ and\ \bibinfo {author} {\bibfnamefont {Y.~R.}\ \bibnamefont {Shen}},\ }\bibfield  {title} {\bibinfo {title} {Optical second-harmonic generation from magnetized surfaces},\ }\href@noop {} {\bibfield  {journal} {\bibinfo  {journal} {Phys.\ Rev.\ B}\ }\textbf {\bibinfo {volume} {39}},\ \bibinfo {pages} {1229} (\bibinfo {year} {1989})}\BibitemShut {NoStop}%
\bibitem [{\citenamefont {Murzina}\ \emph {et~al.}(2021)\citenamefont {Murzina}, \citenamefont {Radovskaya}, \citenamefont {Pashen'kin}, \citenamefont {Gusev}, \citenamefont {Maydykovskiy},\ and\ \citenamefont {Mamonov}}]{Murzina_OptExpr}%
  \BibitemOpen
  \bibfield  {author} {\bibinfo {author} {\bibfnamefont {T.~V.}\ \bibnamefont {Murzina}}, \bibinfo {author} {\bibfnamefont {V.~V.}\ \bibnamefont {Radovskaya}}, \bibinfo {author} {\bibfnamefont {I.~Y.}\ \bibnamefont {Pashen'kin}}, \bibinfo {author} {\bibfnamefont {N.~S.}\ \bibnamefont {Gusev}}, \bibinfo {author} {\bibfnamefont {A.~I.}\ \bibnamefont {Maydykovskiy}},\ and\ \bibinfo {author} {\bibfnamefont {E.~A.}\ \bibnamefont {Mamonov}},\ }\bibfield  {title} {\bibinfo {title} {Effect of inhomogeneous magnetization in optical second harmonic generation from layered nanostructures},\ }\href@noop {} {\bibfield  {journal} {\bibinfo  {journal} {Opt. Expr.}\ }\textbf {\bibinfo {volume} {29}},\ \bibinfo {pages} {2106} (\bibinfo {year} {2021})}\BibitemShut {NoStop}%
\bibitem [{\citenamefont {Kolmychek}\ \emph {et~al.}(2020)\citenamefont {Kolmychek}, \citenamefont {Radovskaya}, \citenamefont {Lazareva}, \citenamefont {Shalygina}, \citenamefont {Gusev}, \citenamefont {Maidykovskii},\ and\ \citenamefont {Murzina}}]{Kolmychek_JETP}%
  \BibitemOpen
  \bibfield  {author} {\bibinfo {author} {\bibfnamefont {I.~A.}\ \bibnamefont {Kolmychek}}, \bibinfo {author} {\bibfnamefont {V.~V.}\ \bibnamefont {Radovskaya}}, \bibinfo {author} {\bibfnamefont {K.~A.}\ \bibnamefont {Lazareva}}, \bibinfo {author} {\bibfnamefont {E.~E.}\ \bibnamefont {Shalygina}}, \bibinfo {author} {\bibfnamefont {N.~S.}\ \bibnamefont {Gusev}}, \bibinfo {author} {\bibfnamefont {A.~I.}\ \bibnamefont {Maidykovskii}},\ and\ \bibinfo {author} {\bibfnamefont {T.~V.}\ \bibnamefont {Murzina}},\ }\bibfield  {title} {\bibinfo {title} {Magnetic-field-induced optical second-harmonic generation study of co/pt and co/ta interfaces},\ }\href@noop {} {\bibfield  {journal} {\bibinfo  {journal} {JETP}\ }\textbf {\bibinfo {volume} {130}},\ \bibinfo {pages} {555} (\bibinfo {year} {2020})}\BibitemShut {NoStop}%
\bibitem [{\citenamefont {Murzina}\ \emph {et~al.}(2020)\citenamefont {Murzina}, \citenamefont {Kolmychek}, \citenamefont {Gusev},\ and\ \citenamefont {Maidykovskii}}]{Murzina_JETPL}%
  \BibitemOpen
  \bibfield  {author} {\bibinfo {author} {\bibfnamefont {T.~V.}\ \bibnamefont {Murzina}}, \bibinfo {author} {\bibfnamefont {I.~A.}\ \bibnamefont {Kolmychek}}, \bibinfo {author} {\bibfnamefont {N.~S.}\ \bibnamefont {Gusev}},\ and\ \bibinfo {author} {\bibfnamefont {A.~I.}\ \bibnamefont {Maidykovskii}},\ }\bibfield  {title} {\bibinfo {title} {Giant magnetic field induced effects in the second-harmonic generation in a planar anisotropic ta/co/pt structure},\ }\href@noop {} {\bibfield  {journal} {\bibinfo  {journal} {JETP Letters}\ }\textbf {\bibinfo {volume} {111}},\ \bibinfo {pages} {333} (\bibinfo {year} {2020})}\BibitemShut {NoStop}%
\bibitem [{\citenamefont {Tserkovnyak}\ \emph {et~al.}(2002)\citenamefont {Tserkovnyak}, \citenamefont {Brataas},\ and\ \citenamefont {Bauer}}]{Tserkovnyak_PRL}%
  \BibitemOpen
  \bibfield  {author} {\bibinfo {author} {\bibfnamefont {Y.}~\bibnamefont {Tserkovnyak}}, \bibinfo {author} {\bibfnamefont {A.}~\bibnamefont {Brataas}},\ and\ \bibinfo {author} {\bibfnamefont {G.~E.~W.}\ \bibnamefont {Bauer}},\ }\bibfield  {title} {\bibinfo {title} {Enhanced gilbert damping in thin ferromagnetic films},\ }\href@noop {} {\bibfield  {journal} {\bibinfo  {journal} {Phys.\ Rev.\ Lett.}\ }\textbf {\bibinfo {volume} {88}},\ \bibinfo {pages} {117601} (\bibinfo {year} {2002})}\BibitemShut {NoStop}%
\bibitem [{\citenamefont {Tserkovnyak}\ \emph {et~al.}(2005)\citenamefont {Tserkovnyak}, \citenamefont {Brataas}, \citenamefont {Bauer},\ and\ \citenamefont {Halperin}}]{Tserkovnyak_RMP}%
  \BibitemOpen
  \bibfield  {author} {\bibinfo {author} {\bibfnamefont {Y.}~\bibnamefont {Tserkovnyak}}, \bibinfo {author} {\bibfnamefont {A.}~\bibnamefont {Brataas}}, \bibinfo {author} {\bibfnamefont {G.~E.~W.}\ \bibnamefont {Bauer}},\ and\ \bibinfo {author} {\bibfnamefont {B.~I.}\ \bibnamefont {Halperin}},\ }\bibfield  {title} {\bibinfo {title} {Nonlocal magnetization dynamics in ferromagnetic heterostructures},\ }\href@noop {} {\bibfield  {journal} {\bibinfo  {journal} {Rev.\ Mod.\ Phys.}\ }\textbf {\bibinfo {volume} {77}},\ \bibinfo {pages} {1375} (\bibinfo {year} {2005})}\BibitemShut {NoStop}%
\bibitem [{\citenamefont {Kapitza}(1951)}]{Kapitza}%
  \BibitemOpen
  \bibfield  {author} {\bibinfo {author} {\bibfnamefont {P.~L.}\ \bibnamefont {Kapitza}},\ }\bibfield  {title} {\bibinfo {title} {Pendulum with a vibrating suspension},\ }\href@noop {} {\bibfield  {journal} {\bibinfo  {journal} {Usp. Fiz. Nauk}\ }\textbf {\bibinfo {volume} {44}},\ \bibinfo {pages} {7} (\bibinfo {year} {1951})}\BibitemShut {NoStop}%
\bibitem [{\citenamefont {Akhiezer}\ and\ \citenamefont {Peletminskii}(1968)}]{Akhiezer_FTT}%
  \BibitemOpen
  \bibfield  {author} {\bibinfo {author} {\bibfnamefont {A.~I.}\ \bibnamefont {Akhiezer}}\ and\ \bibinfo {author} {\bibfnamefont {S.~V.}\ \bibnamefont {Peletminskii}},\ }\href@noop {} {\bibfield  {journal} {\bibinfo  {journal} {Sov. Phys. Solid State}\ }\textbf {\bibinfo {volume} {10}},\ \bibinfo {pages} {2609} (\bibinfo {year} {1968})}\BibitemShut {NoStop}%
\bibitem [{\citenamefont {Zvezdin}\ and\ \citenamefont {Red’ko}(1975)}]{Zvezdin_JETPL}%
  \BibitemOpen
  \bibfield  {author} {\bibinfo {author} {\bibfnamefont {A.~K.}\ \bibnamefont {Zvezdin}}\ and\ \bibinfo {author} {\bibfnamefont {V.~G.}\ \bibnamefont {Red’ko}},\ }\href@noop {} {\bibfield  {journal} {\bibinfo  {journal} {JETP Lett.}\ }\textbf {\bibinfo {volume} {21}},\ \bibinfo {pages} {203} (\bibinfo {year} {1975})}\BibitemShut {NoStop}%
\bibitem [{\citenamefont {Dzhezherya}\ \emph {et~al.}(2012)\citenamefont {Dzhezherya}, \citenamefont {Demishev},\ and\ \citenamefont {Korenivskii}}]{Dzhezherya_JETP}%
  \BibitemOpen
  \bibfield  {author} {\bibinfo {author} {\bibfnamefont {Y.~I.}\ \bibnamefont {Dzhezherya}}, \bibinfo {author} {\bibfnamefont {K.~O.}\ \bibnamefont {Demishev}},\ and\ \bibinfo {author} {\bibfnamefont {V.~N.}\ \bibnamefont {Korenivskii}},\ }\bibfield  {title} {\bibinfo {title} {Kapitza problem for the magnetic moments of synthetic antiferromagnetic syste},\ }\href@noop {} {\bibfield  {journal} {\bibinfo  {journal} {JETP Lett.}\ }\textbf {\bibinfo {volume} {115}},\ \bibinfo {pages} {284} (\bibinfo {year} {2012})}\BibitemShut {NoStop}%
\bibitem [{\citenamefont {Kulikov}\ \emph {et~al.}(2022)\citenamefont {Kulikov}, \citenamefont {Anghel}, \citenamefont {Preda}, \citenamefont {Nashaat}, \citenamefont {Sameh},\ and\ \citenamefont {Shukrinov}}]{Shukrinov_PRB}%
  \BibitemOpen
  \bibfield  {author} {\bibinfo {author} {\bibfnamefont {K.~V.}\ \bibnamefont {Kulikov}}, \bibinfo {author} {\bibfnamefont {D.~V.}\ \bibnamefont {Anghel}}, \bibinfo {author} {\bibfnamefont {A.~T.}\ \bibnamefont {Preda}}, \bibinfo {author} {\bibfnamefont {M.}~\bibnamefont {Nashaat}}, \bibinfo {author} {\bibfnamefont {M.}~\bibnamefont {Sameh}},\ and\ \bibinfo {author} {\bibfnamefont {Y.~M.}\ \bibnamefont {Shukrinov}},\ }\bibfield  {title} {\bibinfo {title} {Kapitza pendulum effects in a josephson junction coupled to a nanomagnet under external periodic drive},\ }\href {https://doi.org/10.1103/PhysRevB.105.094421} {\bibfield  {journal} {\bibinfo  {journal} {Phys. Rev. B}\ }\textbf {\bibinfo {volume} {105}},\ \bibinfo {pages} {094421} (\bibinfo {year} {2022})}\BibitemShut {NoStop}%
\bibitem [{\citenamefont {Landau}\ and\ \citenamefont {Lifshitz}(1984)}]{Landau8}%
  \BibitemOpen
  \bibfield  {author} {\bibinfo {author} {\bibfnamefont {L.~D.}\ \bibnamefont {Landau}}\ and\ \bibinfo {author} {\bibfnamefont {E.~M.}\ \bibnamefont {Lifshitz}},\ }\href@noop {} {\emph {\bibinfo {title} {Course of Theoretical Physics, Vol. 8: Electrodynamics of Continuous Media}}}\ (\bibinfo  {publisher} {Butterworth–Heinemann, Oxford},\ \bibinfo {year} {1984})\BibitemShut {NoStop}%
\bibitem [{\citenamefont {Gaponov}\ and\ \citenamefont {Miller}(1958)}]{Gaponov_JETP}%
  \BibitemOpen
  \bibfield  {author} {\bibinfo {author} {\bibfnamefont {A.~V.}\ \bibnamefont {Gaponov}}\ and\ \bibinfo {author} {\bibfnamefont {M.~A.}\ \bibnamefont {Miller}},\ }\bibfield  {title} {\bibinfo {title} {Potential wells for charged particles in a high-frequency electromagnetic field},\ }\href@noop {} {\bibfield  {journal} {\bibinfo  {journal} {J. Exptl. Theoret. Phys. (U.S.S.R.)}\ }\textbf {\bibinfo {volume} {34}},\ \bibinfo {pages} {242} (\bibinfo {year} {1958})}\BibitemShut {NoStop}%
\bibitem [{\citenamefont {Sinova}\ \emph {et~al.}(2015)\citenamefont {Sinova}, \citenamefont {ands J.~Wunderlich}, \citenamefont {Back},\ and\ \citenamefont {Jungwirth}}]{Sinova_RMP}%
  \BibitemOpen
  \bibfield  {author} {\bibinfo {author} {\bibfnamefont {J.}~\bibnamefont {Sinova}}, \bibinfo {author} {\bibfnamefont {S.~O.~V.}\ \bibnamefont {ands J.~Wunderlich}}, \bibinfo {author} {\bibfnamefont {C.~H.}\ \bibnamefont {Back}},\ and\ \bibinfo {author} {\bibfnamefont {T.}~\bibnamefont {Jungwirth}},\ }\bibfield  {title} {\bibinfo {title} {Spin hall effects},\ }\href@noop {} {\bibfield  {journal} {\bibinfo  {journal} {Rev.\ Mod.\ Phys.}\ }\textbf {\bibinfo {volume} {87}},\ \bibinfo {pages} {1213} (\bibinfo {year} {2015})}\BibitemShut {NoStop}%
\bibitem [{\citenamefont {Rzhevsky}\ \emph {et~al.}(2007)\citenamefont {Rzhevsky}, \citenamefont {Krichevtsov}, \citenamefont {B{\"u}rgler},\ and\ \citenamefont {Schneider}}]{Rzhevsky_PRB}%
  \BibitemOpen
  \bibfield  {author} {\bibinfo {author} {\bibfnamefont {A.~A.}\ \bibnamefont {Rzhevsky}}, \bibinfo {author} {\bibfnamefont {B.~B.}\ \bibnamefont {Krichevtsov}}, \bibinfo {author} {\bibfnamefont {D.~E.}\ \bibnamefont {B{\"u}rgler}},\ and\ \bibinfo {author} {\bibfnamefont {C.~M.}\ \bibnamefont {Schneider}},\ }\bibfield  {title} {\bibinfo {title} {Interfacial magnetization in exchange-coupled fe/cr/fe structures investigated by second harmonic generation},\ }\href@noop {} {\bibfield  {journal} {\bibinfo  {journal} {Phys.\ Rev.\ B}\ }\textbf {\bibinfo {volume} {75}},\ \bibinfo {pages} {144416} (\bibinfo {year} {2007})}\BibitemShut {NoStop}%
\bibitem [{\citenamefont {Yu}\ \emph {et~al.}(2014)\citenamefont {Yu}, \citenamefont {d'Allivy Kelly}, \citenamefont {Cros}, \citenamefont {Bernard}, \citenamefont {Bortolotti}, \citenamefont {Anane}, \citenamefont {Brandl}, \citenamefont {Huber}, \citenamefont {Stasinopoulos},\ and\ \citenamefont {Grundler}}]{Yu_SciRep}%
  \BibitemOpen
  \bibfield  {author} {\bibinfo {author} {\bibfnamefont {H.}~\bibnamefont {Yu}}, \bibinfo {author} {\bibfnamefont {O.}~\bibnamefont {d'Allivy Kelly}}, \bibinfo {author} {\bibfnamefont {V.}~\bibnamefont {Cros}}, \bibinfo {author} {\bibfnamefont {R.}~\bibnamefont {Bernard}}, \bibinfo {author} {\bibfnamefont {P.}~\bibnamefont {Bortolotti}}, \bibinfo {author} {\bibfnamefont {A.}~\bibnamefont {Anane}}, \bibinfo {author} {\bibfnamefont {F.}~\bibnamefont {Brandl}}, \bibinfo {author} {\bibfnamefont {R.}~\bibnamefont {Huber}}, \bibinfo {author} {\bibfnamefont {I.}~\bibnamefont {Stasinopoulos}},\ and\ \bibinfo {author} {\bibfnamefont {D.}~\bibnamefont {Grundler}},\ }\bibfield  {title} {\bibinfo {title} {Magnetic thin-film insulator with ultra-low spin wave damping for coherent nanomagnonics},\ }\href@noop {} {\bibfield  {journal} {\bibinfo  {journal} {Sci. Rep.}\ }\textbf {\bibinfo {volume} {4}},\ \bibinfo {pages} {6848} (\bibinfo {year} {2014})}\BibitemShut {NoStop}%
\bibitem [{\citenamefont {Zhang}\ \emph {et~al.}(2016)\citenamefont {Zhang}, \citenamefont {Yue}, \citenamefont {Kou}, \citenamefont {Lin}, \citenamefont {Zhai},\ and\ \citenamefont {Zhai}}]{Zhang_AIP}%
  \BibitemOpen
  \bibfield  {author} {\bibinfo {author} {\bibfnamefont {D.}~\bibnamefont {Zhang}}, \bibinfo {author} {\bibfnamefont {J.~J.}\ \bibnamefont {Yue}}, \bibinfo {author} {\bibfnamefont {Z.~X.}\ \bibnamefont {Kou}}, \bibinfo {author} {\bibfnamefont {L.}~\bibnamefont {Lin}}, \bibinfo {author} {\bibfnamefont {Y.}~\bibnamefont {Zhai}},\ and\ \bibinfo {author} {\bibfnamefont {H.~R.}\ \bibnamefont {Zhai}},\ }\bibfield  {title} {\bibinfo {title} {The investigation of ferromagnetic resonance linewidth in ni80fe20 films with submicron rectangular elements},\ }\href@noop {} {\bibfield  {journal} {\bibinfo  {journal} {AIP Advances}\ }\textbf {\bibinfo {volume} {6}},\ \bibinfo {pages} {056125} (\bibinfo {year} {2016})}\BibitemShut {NoStop}%
\bibitem [{\citenamefont {Krivulin}\ \emph {et~al.}(2023)\citenamefont {Krivulin}, \citenamefont {Pashen'kin}, \citenamefont {Gorev}, \citenamefont {Yunin}, \citenamefont {Sapozhnikov}, \citenamefont {Grunin}, \citenamefont {Zaharova},\ and\ \citenamefont {Leont'yev}}]{Dorofey_JTP}%
  \BibitemOpen
  \bibfield  {author} {\bibinfo {author} {\bibfnamefont {D.~O.}\ \bibnamefont {Krivulin}}, \bibinfo {author} {\bibfnamefont {I.~Y.}\ \bibnamefont {Pashen'kin}}, \bibinfo {author} {\bibfnamefont {R.~V.}\ \bibnamefont {Gorev}}, \bibinfo {author} {\bibfnamefont {P.~A.}\ \bibnamefont {Yunin}}, \bibinfo {author} {\bibfnamefont {M.~V.}\ \bibnamefont {Sapozhnikov}}, \bibinfo {author} {\bibfnamefont {A.~V.}\ \bibnamefont {Grunin}}, \bibinfo {author} {\bibfnamefont {S.~A.}\ \bibnamefont {Zaharova}},\ and\ \bibinfo {author} {\bibfnamefont {V.~N.}\ \bibnamefont {Leont'yev}},\ }\href@noop {} {\bibfield  {journal} {\bibinfo  {journal} {J. Techn. Phys.}\ }\textbf {\bibinfo {volume} {93}},\ \bibinfo {pages} {907} (\bibinfo {year} {2023})}\BibitemShut {NoStop}%
\bibitem [{\citenamefont {Skorokhodov}\ \emph {et~al.}(2017)\citenamefont {Skorokhodov}, \citenamefont {Demidov}, \citenamefont {Vdovichev},\ and\ \citenamefont {Fraerman}}]{Skorokhodov_JETP}%
  \BibitemOpen
  \bibfield  {author} {\bibinfo {author} {\bibfnamefont {E.~V.}\ \bibnamefont {Skorokhodov}}, \bibinfo {author} {\bibfnamefont {E.~S.}\ \bibnamefont {Demidov}}, \bibinfo {author} {\bibfnamefont {S.~N.}\ \bibnamefont {Vdovichev}},\ and\ \bibinfo {author} {\bibfnamefont {A.~A.}\ \bibnamefont {Fraerman}},\ }\bibfield  {title} {\bibinfo {title} {Ferromagnetic resonance in a system of magnetic films with different curie temperatures},\ }\href@noop {} {\bibfield  {journal} {\bibinfo  {journal} {JETP}\ }\textbf {\bibinfo {volume} {124}},\ \bibinfo {pages} {617} (\bibinfo {year} {2017})}\BibitemShut {NoStop}%
\end{thebibliography}%

\end{document}